# Generative AI as a metacognitive agent:
# A comparative mixed-method study with human participants on ICF-mimicking exam performance

Jelena Pavlović
*University of Belgrade, Faculty of Philosophy & Koučing centar Resarch Lab*

Jugoslav Krstić, Luka Mitrović, Đorđe Babić,
Adrijana Milosavljević, Milena Nikolić, Tijana Karaklić & Tijana Mitrović
*Koučing centar Research Lab*

*Abstract*

This study investigates the metacognitive capabilities of Large Language Models (LLMs) relative to human metacognition in the context of the International Coaching Federation (ICF)-mimicking exam, a situational judgment test related to coaching competencies. Using a mixed-method approach, we assessed the metacognitive performance—including sensitivity, accuracy in probabilistic predictions, and bias—of human participants and five advanced LLMs: GPT-4, Claude-3-Opus 3, Mistral Large, Llama 3, and Gemini 1.5 Pro. The results indicate that LLMs outperformed humans across all metacognitive metrics, particularly in terms of reduced overconfidence, compared to humans. However, both LLMs and humans showed less adaptability in ambiguous scenarios, adhering closely to predefined decision frameworks. The study suggests that Generative AI can effectively engage in human-like metacognitive processing without conscious awareness. Implications of the study are discussed in relation to development of AI simulators that scaffold cognitive and metacognitive aspects of mastering coaching competencies. More broadly, implications of these results are discussed in relation to development of metacognitive modules that lead towards more autonomous and intuitive AI systems.

## Introduction

Metacognition, the ability to understand and regulate one's cognitive processes, is a fundamental aspect of human learning, decision making and problem solving. Traditionally viewed as a conscious process, metacognition involves activities such as planning, monitoring, and evaluating one's performance during cognitive tasks. However, recent studies suggest that certain metacognitive processes can occur without conscious awareness, challenging the traditional boundaries of how metacognition is understood and measured Kentridge and Heywood (2000).

In the field of generative artificial intelligence, particularly in Large Language Models (LLMs), metacognitive-like processes may manifest as algorithms adapt, learn, and optimize performance. This raises intriguing questions about the nature of metacognition in non-conscious entities and its comparison to human metacognitive processes. The present study aims to explore these questions by comparing the metacognitive processes of human participants and LLMs within the context of the International Coaching Federation (ICF) exam performance. By examining metacognitive patterns of exam performance, this research seeks to gain insights into the similarities and differences in how metacognition works for human and artificial intelligences.

## (Human) Metacognition

Metacognition is the term introduced to describe "cognition about cognitive phenomena" (Flavell, 1979). The phenomenon was introduced to point to the importance of monitoring over human cognitive processes, especially in the developmental and educational perspective. According to Flavell (1979), there are four classes of metacognitive phenomena: (1) metacognitive knowledge; (2) metacognitive experiences; (3) goal setting; (4) strategy deployment. Example of a metacognitive knowledge would be our awareness of our cognitive style, while metacognitive experience could refer to a feeling of enjoyment in solving a certain task at hand. We use our metacognitive knowledge and experience to set goals and employ certain strategies of reaching those goals. Akturk & Sahin (2011) have outlined an important distinction between cognitive and metacognitive strategies. While cognitive strategies include, for example, summarizing or conceptual mapping, metacognitive strategies involve planning, goal setting, monitoring and evaluating cognitive performance.

Metacognition plays a pivotal role in test performance (Sulaiman et al., 2022). It enables persons to plan, monitor, and assess their test understanding and performance. Metacognitive skills are particularly important in types of test performance which requires creating and adapting strategies (e.g. allocating time, elimination methods, logical deduction, use of heuristics) which contribute to optimizing test performance. Swanson (1990) found that metacognitive knowledge may compensate for lower overall aptitude in problem solving. This study found that higher metacognitive ability groups were more likely to rely on application of specific frameworks (hypothetico-deductive strategies) and evaluation strategies than the lower metacognitive group (Swanson, 1990). Importance of metacognition in test performance points out the possible role of instruction or training in metacognitive skills to better prepare test takers. Important aspect of the metacognition research paradigm has remained a belief in its developmental potential and positive impact on the learning process. Studies have revealed that metacognition develops around preschool and early school years, initially in separate domains, while later becoming generalized with the role of feedback and metacognitive instruction (Veenman, 2005; Veenman & Spaans, 2005).

**Metacognitive metrics**

Gaining insight into human metacognition included the need to ask participants about their cognitions, usually during a certain task performance. Veenman (2005) noted the use of questionnaires, interviews, thinking-aloud-protocols, stimulated recall observations and many others. Over time there has been a growing number of metacognitive metrics (Fleming & Lau, 2014).

*Absolute accuracy* refers to proportion of correct answers on a test, representing a measure of cognitive, rather than metacognitive performance. A high absolute accuracy indicates that the individual is performing well in terms of tasks. However, it does not provide information about how well an individual is monitoring and evaluating his or her performance. As a hypothetical example, an individual may score 65% on a test, responding to cognitive performance. If we ask how confident an individual is about his or her response, we are entering the field of metacognition as judgment of own cognitive performance.

*Metacognitive sensitivity* assesses how well an individual can distinguish between his or her correct and incorrect responses. If an individual has a high hit rate (e.g. confidently identifies many of correct answers) and a low false alarm rate (e.g. doesn't mistakenly claim high confidence in many incorrect answers), this would signify good metacognitive sensitivity.

If an individual generally tends to overestimate or underestimate own performance, we may point to *metacognitive bias*. Bias in the context of metacognitive performance refers to systematic errors in confidence judgments. Positive bias indicates overconfidence, where confidence ratings are higher than warranted by actual correctness. Understanding bias provides insights into how individuals perceive and evaluate their own performance relative to their actual abilities.

The Brier score is used as a metric that assesses the *accuracy of probabilistic predictions* by comparing predicted probabilities (confidence ratings) with actual outcomes (correctness). A lower Brier score indicates better calibration of confidence judgments, meaning that confidence ratings are more accurate in predicting correctness. A higher Brier score suggests poorer calibration, indicating that confidence ratings are less reliable in reflecting actual performance. While the Brier score evaluates the precision and reliability of probability estimates in predictions, bias assesses whether individuals have an accurate self-assessment of their capabilities or knowledge as reflected in their confidence levels.

**Metacognition and consciousness**

Another line of research of importance to this study is study into metacognition and consciousness. Kentridge & Heywood (2000) have carried out experiments which showed that learning novel cognitive schemas may not involve metacognition, while adaptive learning does involve metacognition. Again, taking the exam performance as an example: a person may score high on a test, which is not in itself an indicator of metacognition. But if a person changes exam performance based on previous performance, we may say that metacognition is involved. Moreover, these authors propose that it is possible that metacognitive processes operate without an individual being aware of adaptive learning. In other words, we may say that metacognition is not an indicator of consciousness in itself.

In connection to consciousness debates, Fleming & Lau (2014) argue that low metacognitive sensitivity may indicate absence of conscious experience, but may also be an indicator of hallucinations in clinical contexts. In humans, high discrepancies between actual performance and confidence levels may be a predictor of hallucinations (Wright et al., 2024). That is why interventions that target improvements in metacognition may serve as hallucination control tools. What we are learning from these studies is that low metacognitive sensitivity may be indicative of absence or distortions in consciousness.

**Artificial metacognition?**

We already know that Large Language Models (LLMs) successfully perform on a variety of traditionally human exams, from biomedical science (Stribling et al., 2024) to emotional intelligence (Wang et al., 2023). Furthermore, in the rapidly evolving field of generative artificial intelligence, metacognitive-like processes are increasingly observed. Current research has tested possibility of creation of metacognition modules of LLMs in game simulations, pointing to their ability to pick up new insights, draw conclusions from past memories and use these insights for future actions (Toy et al., 2024). Studies have illustrated how LLMs impose metacognitive demands on users and proposed addressing this by integrating metacognitive support strategies or scaffolding meta-prompts into LLM design (Li et al., 2023; Suzgun & Kalai, 2024; Tankelevitch et al., 2024). A tendency of LLM towards overconfidence has been outlined with proposed prompting strategies to mitigate overconfidence (Xiong, 2024).

Unlike humans, however, these models operate without conscious experience, raising further fundamental questions about the nature and necessity of consciousness in metacognitive functions. We may assume that LLMs exhibit behaviors that suggest a form of "artificial metacognition." For instance, these models could assign certainty levels to their exam responses or explain their judgment of certainty levels. They may adjust their output strategies based on the feedback received from humans. This adaptive behavior is crucial for tasks in which continual adjustment is paramount. Studies pointing to metacognitive abilities of LLMs have served the supportive role in developing the concept of generative AI agents, which can elicit and orchestrate different functional strengths of LLMs. Continuous improvements in generative AI as a metacognitive agent seems as a promising pathway for both research and practice.

## Research gap and problem statement

Comparison between human and AI metacognitive processes is critical for several reasons. Firstly, it can help demystify whether AI can truly mimic human metacognitive processes or if these processes merely simulate superficial aspects of human metacognition. For example, it is interesting to explore to what extent AI outputs are result of processing that we would classify as cognitive or metacognitive. Even to prompts that aim to elicit metacognitive processing, we still may obtain only cognitive processing by AI. Secondly, understanding these differences and similarities can inform the development of more intuitive and effective AI systems, which are better aligned with human metacognition. Lastly, such comparative studies can provide deeper insights into the essential elements of consciousness and awareness in metacognition, potentially deepening our understanding of these processes in both biological and artificial entities. While LLMs have shown proficiency in various cognitive tasks, their ability to engage in metacognitive processes, that involve monitoring and regulating one's own thought processes, remains underexplored.

Metacognitive skills are particularly important for optimizing performance in complex tests, such as situational judgment tests (SJT), which present test takers with prototypical work-related problems followed by a set of possible solutions (Rasmussen, 2009). Among many types of SJTs, those applied to complex professional domains seem to be most plausible for exploring metacognitive processes.

Coaching is a profession grounded in self-awareness and the capacity for reflective practice. Coaching exams often test these attributes, providing insights into how individuals use metacognition to evaluate competencies, beliefs and values in a professional context. We assumed that this type of test engages metacognitive processes as candidates must appraise each option and justify their selected response. For the purpose of this study a SJT in the coaching domain was chosen as a type of exam rich in the cognitive and metacognitive demands it places on candidates, making it an important and relevant tool for studying metacognition.

Aims of the study were to assess and compare metacognitive performance of human participants and LLMS in their their ability to select the best and worst answers during a SJT in a coaching domain, as well as to compare justification of judgment used to explain their answers.

## Method

### Participants

*Human Participants*

The study involved total of 3 human participants. One human subject matter expert (SME) was trained as ICF assessor, serving as a benchmark for correct test performance. Other 2 human participants served the role of test takers. Participants were selected based on their professional experience in fields relevant to the ICF exam content. Test takers were chosen based on the criterion that they actually passed the official ICF exam within the previous 3 months. Participants were recruited through professional networks, ensuring a diverse representation of expertise levels and backgrounds.

*Artificial Participants*

Five advanced LLMs, specifically designed for complex problem-solving and natural language processing tasks, were used. These models included the latest versions of GPT4, Gemini 1.5 pro, Llama 3, Claude-3-Opus 3 and Mistral Large.

**Materials**

*ICF-mimicking exam content*

Human subject matter expert co-created a set of 50 multiple-choice questions in collaborative work with GPT4 (Example in Table 1). Out of this question set, 20 questions were randomly taken for the purpose of the study. In line with the ICF format of situational judgement tests, the questions required that participants evaluate multiple potential responses to a single scenario ("best" and "worst" responses). Exam content was derived from sample of 8 sample questions that are publicly available on the ICF web page. These covered a range of topics relevant to coaching methodologies and ethical considerations.

*Table 1. Example of an ICF-mimicking test question.*

> 1. Question: A client is happy about a recent promotion but anxious about the new responsibilities. What should the coach do?
> - A: Suggest the client to focus only on the positive aspects of the promotion and ignore the anxiety.
> - B: Encourage the client to express their feelings about the promotion and the anxiety, exploring both aspects.
> - C: Tell the client that they just need to work harder to overcome their anxiety.
> - D: Recommend the client seek therapy if they feel anxious about their job responsibilities.
> 
> What is the best option?
> What is the worst option?

Each question had predefined best and worst answers, established by the subject matter expert who participated in the study.

**Procedures**

*Task execution and exam administration*

Both human participants and the LLMs completed the ICF-mimicking exam. They were asked to identify the best and worst answers for each question and rate their confidence in each choice on a scale from 0 to 100%, providing a brief explanation of their response. All participants processed the same set of questions. Instruction was as follows: "You will be presented with a set of questions for the ICF-mimicking exam. Your task is to (1) Go through the questions and possible solutions (best and worst responses); 2) pick one "best" and one "worst" solution in line with the ICF Core competencies; 3) evaluate each "best" and "worst" response on a scale 1-100 according to the certainty level (how certain you are that the response is correct); (2) Now reflect and explain why you assigned certain levels of confidence to each question in line with the ICF competencies. Try to summarize your reasoning process into 2-3 sentences. Take a question by question and reflect on your reasoning case by case."

### Data Collection

For human participants, responses were collected via a digital interface that logs choices, confidence levels and explanations of judgement. For LLMs, logs of their decision-making processes, including confidence scores and explanations of reasoning were extracted directly from the model's output data.

### Data Analysis

Comparative mixed-method (quantitative and qualitative) analysis was conducted to account for metacognitive patterns in human and LLM participants.

### Statistical analysis

Quantitative measures of metacognitive performance included in the study for comparative analysis between human participants and LLM are described below.

*Metacognitive sensitivity, a*s an ability to discriminate between correct and incorrect responses based on confidence rating, was calculated using the following formula:
*Sensitivity=High Confidence Hit Rate−High Confidence False Alarm Rate.*

*Accuracy of probabilistic predictions*. The Brier score was used as a type of metacognitive metric that measures the accuracy of probabilistic predictions using this formula:
*Brier Score=N1∑(Predicted Probability−Actual Outcome)2*

*Metacognitive Bias,* as a tendency towards overconfidence or under confidence, was evaluated by using this formula: *Metacognitive Bias=Average Confidence−Proportion of Correct Responses*

### Qualitative analysis

Inductive qualitative thematic analysis was performed to gain insight into the patterns of reasoning justification used by human participants and LLMs. The coding scheme that emerged during the analysis is as presented in the Table 2. Once the coding scheme was created, a frequency analysis was carried out to ground qualitative claims in data.

*Table 2. Coding scheme for justification of judgment responses.*

| Category | Example |
|---|---|
| Coach behavior | *Facilitates a comprehensive evaluation of the decision by considering various factors and potential consequences. (Gemini 1.5 Pro Pro)* |
| Client outcome | *Best response aids in effective management skills (GPT 4)* |
| ICF competencies not connected to coach behavior | *Best response aligns with ICF competencies of promoting awareness and facilitating client growth. (GPT 4)* |
| ICF competencies connected to coach behavior | *This response aligns with the ICF competency of "Establishing trust and intimacy with the client" by exploiting the client's underlying concerns and fears. (Llama3)* |
| Reflective comments | *Inviting client to explore the meaning of the failure would be right thing to do. I'm not sure I would invite them to think of worst case scenario here. (Human Subject Matter Expert)* |

## Results

**Quantitative analysis**

In this section, data on metacognitive metrics of humans and large language models (LLMs), focusing on ICF-mimicking exam are presented. Results on metacognitive metrics for human and LLM participants are organized in the following manner: 1. Analysis of "best" responses for human participants and LLM participants as a group, with additional analysis for specific LLMs; 2. Analysis of "worst" responses for human participants and LLM participants as a group, with additional analysis for specific LLMs.

*"Best" Responses: Human participants compared to LLMS*

Both humans and LLMs achieved perfect scores on "Best" responses (100% for both participant groups), indicating that this type of task was too easy and indiscriminative of cognitive performance. Overall data for "best" responses (Table 3) point that both humans and LLMs demonstrate excellent metacognitive metrics. Both groups have almost perfect discrimination ability in their best responses. Both groups show zero or negative bias, with humans at perfect zero and LLMs exhibiting a low average negative bias, indicating a slightly conservative estimation of their performance capabilities.

*Table 3. Metacognitive metrics for "Best Responses"
of human participants and LLM participants.*

| Metric | Humans | LLMs |
|---|---|---|
| Metacognitive sensitivity | 1.00 | 0.98 |
| Brier Score | 0.00 | 0.00 |
| Metacognitive Bias | 0.00 | -0.07 |

## "Best" responses data: Comparing specific LLMs

Metacognitive metrics for "best" responses are similar among LLMs (Table 4). All models except Gemini 1.5 Pro show high sensitivity by assigning high confidence only to correct answers and avoiding false alarms. Low Brier Scores suggest accurate probability calibration among the models. Negative values of bias indicate a tendency of slight under confidence of all LLMs, despite accuracy when choosing "best" responses.

Table 4. Metacognitive metrics for "Best Responses" of specific LLM participants.

| LLM | Metacognitive sensitivity | Brier Score | Metacognitive Bias |
|---|---|---|---|
| GPT 4 | 1.00 | 0.00 | -0.05 |
| Llama 3 | 1.00 | 0.00 | -0.09 |
| Claude-3-Opus Opus | 1.00 | 0.00 | -0.05 |
| Mistral Large Large | 1.00 | 0.00 | -0.07 |
| Gemini 1.5 Pro pro | 0.90 | 0.01 | -0.10 |

## "Worst" Responses: Human participants compared to LLMS

Overall, both groups exhibited cognitive and metacognitive challenges in judging their "worst" responses (Table 5). LLMs showed higher absolute accuracy (40%) compared to humans (30%), suggesting that LLMs handle difficult scenarios slightly better.

Table 5. Metacognitive Metrics for "Worst Responses" - Humans vs. LLMs.

| Metric | Humans | LLMs |
|---|---|---|
| Metacognitive sensitivity | -0.04 | 0.03 |
| Brier Score | 0.67 | 0.40 |
| Metacognitive Bias | 0.74 | 0.21 |

Both human participants and LLMs demonstrated low metacognitive sensitivity in terms of limited discrimination ability when it comes to judging correct and incorrect "worst" response. LLMs as a group showed moderate Brier score, indicating moderate predictive accuracy, while Brier score of human participants pointed to low predictive accuracy. In terms of bias, human participants tended to have greater levels of over confidence.

Overall, both groups demonstrated metacognitive challenges when confronted with more demanding tasks, with LLMs outperforming human participants in all metacognitive metrics. As a group, LLMs showed low sensitivity, moderate predictive accuracy and moderate bias. Human participants as a group showed negative sensitivity (more incorrect high confidence responses than correct ones), very low predictive accuracy and high over confidence. Metacognitive bias stands out as a key pattern in distinguishing between human participants and LLMs when it comes to performance on difficult tasks.

### Worst" Responses: Comparing specific LLMs

As displayed in Table 6, metacognitive metrics for "worst" responses differ by specific LLMs. GPT 4 showed the highest accuracy (50%), indicating it handles adverse scenarios relatively better than other LLMs. Claude-3-Opus Opus had the lowest accuracy (30%), which might suggest specific weaknesses in more challenging or complex conditions.

*Table 6. Metacognitive Metrics for "Worst Responses" by specific LLMs.*

| LLM | Metacognitive sensitivity | Brier score | Bias |
|---|---|---|---|
| GPT 4 | 0.20 | 0.37 | 0.37 |
| Llama 3 | 0.00 | 0.28 | -0.23 |
| Claude-3-Opus Opus | 0.00 | 0.59 | 0.64 |
| Mistral Large | 0.00 | 0.24 | -0.20 |
| Gemini 1.5 Pro pro | -0.04 | 0.48 | 0.49 |

Overall metacognitive sensitivity varied across the models, with GPT 4 moderately distinguishing correct responses, while assigning high confidence to some incorrect answers. Llama 3, Claude-3-Opus and Mistral Large had low sensitivity, while Gemini 1.5 Pro pro hade higher false alarm rate than hit rate. Mistral Large and Llama 3 showed moderate to low levels of Brier score, indicating somewhat accurate predictions about their judgment about "worst" responses. In terms of predictive accuracy, GPT 4 showed moderate levels, Gemini 1.5 Pro moderate to low, with

Claude-3-Opus Opus demonstrating low levels. Positive values of bias suggest a trend towards overconfidence in "worst" scenarios, with the exception of Llama 3 and Mistral Large, who remain underconfident when facing challenging situations. GPT 4 tended to exhibit moderate overconfidence, Gemini 1.5 Pro moderate to high, and Claude-3-Opus being significantly overconfident.

In summary, when facing challenging tasks, GPT 4 tends to have moderate overall metacognitive metrics (sensitivity, predictive accuracy and bias). Metacognitive pattern of Llama 3 and Mistral Large in context of challenging tasks highly resemble each other, with a marked pattern of under confidence, low sensitivity and moderate predictive accuracy. Gemini 1.5 Pro pro exhibits negative sensitivity, moderate to low predictive accuracy and moderate to high overconfidence. Finally, Claude-3-Opus Opus stands out as a model with a specific metacognitive profile: low sensitivity, low predictive accuracy and high over confidence.

## Qualitative analysis

### Qualitative analysis of justification of reasoning: "Best" responses

Qualitative analysis revealed patterns of responses that differentiate between LLMs's, human participants' and experts' justification of "best" responses (Table 7). In general, categories related to application of ICF competencies (whether connected or not connected to coach behavior) and reflection were categorized as higher metacognitive abilities (Swanson, 1990).

*Table 7. Frequencies of categories found in participants' justifications of reasoning behind the choice of "best" response on the ICF-mimicking exam.*

|  | Coach behavior | Client outcome | ICF competencies not connected to coach behavior | ICF competencies connected to coach behavior | Reflective comments |
|---|---|---|---|---|---|
| GPT 4 | 2 | 8 | 10 | 1 | 0 |
| Gemini 1.5 Pro | 14 | 12 | 0 | 8 | 0 |
| Claude-3-Opus | 0 | 0 | 0 | 20 | 0 |
| Mistral Large | 0 | 0 | 18 | 0 | 0 |
| Llama 3 | 0 | 0 | 0 | 20 | 0 |
| Human participant 1 | 0 | 0 | 0 | 5 | 0 |
| Human participant 2 | 0 | 1 | 2 | 17 | 0 |

| Expert | 4 | 5 | 4 | 8 | 9 |

*Justification of reasoning for "best" responses by LLMs*

In general, LLMs showed a stronger focus on ICF competencies, either connected or not connected to coach behavior. This suggests that LLMs were more oriented towards structured frameworks and standards. Reflective comments were absent among the LLMs, indicating a potential limitation in their ability to engage in reflective practice. GPT-4 displayed a tendency to justify reasoning by relying on ICF competencies, without connecting them to coach behavior. Gemini 1.5 Pro leaned heavily on coach behavior and client outcomes. However, it also took into account ICF competencies connected to coach behavior. Mistral Large stood out with a singular focus on ICF competencies not connected to coach behavior, emphasizing the technical standards without grounding them in behavioral data. Llama 3 and Claude-3-Opus Opus, solely focused on competencies connected to coach behavior, suggesting a reasoning pattern that connects reported behavior to an established set of criteria.

*Justification of reasoning for "best" responses by Human participants*

Human participants tended to use ICF competencies grounded in coach behavior as a justification of their reasoning. High number of missing responses for Human participant 1 may indicate a challenge in providing justification. There's a notable absence of reflective comments, suggesting that reflective practice may not be developed.

*Justification of reasoning for "best" responses by Expert*

The Expert demonstrated a balance across all categories with a unique addition of reflective comments. This balance may indicate a well-rounded approach to judgment, with a reflective practice indicating a depth and complexity of justification. Expert's use of reflective comments points to the value of metacognitive skills in professional judgment, potentially differentiating expert coaches from less experienced ones and AI models.

**Qualitative analysis of justification of reasoning: "Worst" responses**

Patterns of reasoning justification of different participant used to explain "worst" answers during the ICF-mimicking exam differ in case of justifying "worst" responses (Table 8).

*Table 8. Frequencies of categories found in participants' justifications of reasoning behind the choice of "worst" response on the ICF-mimicking exam.*

| | Coach behavior | Client outcome | ICF competencies | ICF competencies | Reflective comments |
|---|---|---|---|---|---|

|  |  |  | not connected to coach behavior | connected to coach behavior |  |
|---|---|---|---|---|---|
| GPT 4 | 10 | 12 | 0 | 0 | 0 |
| Gemini 1.5 Pro Pro | 19 | 7 | 0 | 0 | 0 |
| Claude-3-Opus Opus | 17 | 14 | 0 | 0 | 0 |
| Mistral Large | 0 | 0 | 20 | 0 | 0 |
| Llama 3 | 20 | 7 | 0 | 0 | 0 |
| Human participant 1 | 1 | 1 | 0 | 4 | 0 |
| Human participant 2 | 15 | 1 | 0 | 5 | 0 |
| Expert | 6 | 7 | 0 | 14 | 6 |

*Justification of reasoning for "worst" responses by LLMs*

The absence of ICF competencies connected to coach behavior in the "worst" responses is a notable pattern of justification of LLMs overall. This may suggest interpreting ICF competencies in a narrow or rigid manner by not associating negative outcomes with a lack of adherence to ICF standards as directly as they do positive outcomes. The absence of references to ICF competencies may also reflect LLMs' limitations in considering the contextual nuances that expert coaches use to assess when specific coach behaviors (aligned with ICF competencies) are lacking or inappropriate. Mistral Large is the only LLM that remains consistent in justification of "best" and "worst" responses, focusing on ICF competencies not connected to coach behavior.

*Justification of reasoning for "worst" responses by Human participants*

High number of missing responses for Human participant 1 remains consistent across the justification of the "worst" responses. For the "worst" responses, Human Participant 2 shifted focus from using ICF competencies connected to coach behavior to client outcomes and coach behavior, similarly to LLMs. There remains a consistency in absence of reflective comments.

*Justification of reasoning for "worst" responses by Expert*

In contrast to LLMs and human participants, Expert's focus shifted towards ICF competencies connected to coach behavior, when justifying "worst" responses. This shift may indicate that

Expert perceived violation of ICF competencies connected to coach behavior as a primary contributor to "worst" outcomes. Expert remained consistent in using reflective comments when justifying worst responses as another unique feature among the participants.

## Discussion

This study investigated the metacognition of Large Language Models (LLMs) compared to human participants by evaluating their performance on an ICF-mimicking exam. The goal was to assess whether AI can engage in metacognitive activities like estimating certainty of response after providing exam responses in a situational judgment test on understanding coaching competencies. This research goal may be connected to a recent call for a science of metacognition of artificial systems (Fleming, 2023).

AI outperformed human participants in terms of absolute accuracy in identifying the best responses to the exam questions. This demonstrates a strong understanding of the ICF core competencies in optimal scenarios. Previous studies have also found that LLMs achieved above-average Emotional Quotient (EQ) scores , with GPT-4 exceeded 89% of human participants (Wang et al, 2023). LLMs outperformed human participants in identifying worst response too, although both groups showed a decline in cognitive performance when selecting the worst responses. The task to identify worst responses demands understanding of subtleties that distinguish poor choices from plausible but non-optimal ones. This decline in cognitive performance of both groups of participants can be connected to a broader insight that situational judgment tests with two response alternatives ("best" and "worst" responses) tend to have lower internal consistency (Lievens et al., 2008).

When it comes to metacognitive abilities related to ICF-mimicking test, LLMs also tend to slightly outperform human participants. When tasks are very easy or straightforward, both human participants and LLMs show a tendency toward very good metacognitive abilities. However, when confronted with more challenging or complex tasks, both human participants and LLMs demonstrate metacognitive challenges. A pattern in metacognitive profile refers to LLMs' lower tendency towards overconfidence in their judgments of challenging tasks.

This finding can be connected to multiple previous studies which have detected a human tendency toward overconfidence (Kruger & Dunning, 1999; Mazor & Fleming, 2021; Miller & Geraci, 2014). At the same time, previous studies have pointed to over confidence of LLMs, comparing GPT 3, GPT 3.5, Llama 2, Vicuna and GPT 4 (Xion et al. 2024). However, these differences in findings compared to the present study may be explained by improvements in models that have occurred over time.

When it comes to individual LLMs metacognitive profile, the present study points to underconfidence – overconfidence as a key dimension for their discrimination. On one side we found Llama 3 and Mistral Large to be underconfident in judgment, while Claude-3-Opus Opus tends to be highly overconfident. No matter how rapidly LLMs evolve, these findings may serve as data points in present time and benchmarks for further evaluation studies of their improvements.

The qualitative analysis of justifications for "best" and "worst" coaching responses by LLMs, human participants, and an expert demonstrates differences in metacognitive patterns. Human participants and LLMs tend to face challenges in consistently applying the same criteria under uncertainty, while consistently lacking reflective commentary. Expert distinguishes through a balanced and consistent use of justification across contexts with a unique capacity for self-reflection.

The juxtaposition of quantitative and qualitative findings indicates that while LLMs are proficient in exam performance and metacognitive evaluation in structured scenarios, they struggle with more complex judgment tasks that require a nuanced understanding and reflective thinking—areas where human experts excel. The qualitative data underscores the importance of integrating reflective practices into the training of both AI and human professionals to enhance their metacognitive judgment capabilities in complex situations.

Overall, the findings of the present study that LLMs exhibit lower overconfidence in their metacognitive patterns and tend to be more conservative in adverse situations, may sound intriguing. Previous studies have reported that human overconfidence may drive further confirmation bias as selective disregard of counter-evidence (Rollwage et al., 2020). It seems that in human-LLM interaction we may find useful metacognitive scaffolding to help reduce cognitive and metacognitive biases. Findings of the present study can also be seen in light of experimental data by Colombatto & Fleming (2023), who found that people were more willing to accept advice of AI systems compared to humans, indicating a systematic illusion of confidence in AI decisions. The present study confirms the importance of metacognition in guiding human-machine interactions. In this interaction we need to understand our own cognitive biases towards AI and also understand metacognitive functioning and limitations of available AI systems. Interaction with AI strongly depends on continuous understanding and improvement of metacognitive abilities of AI, as well as skillful use of human metacognition to navigate towards informed and ethical collaboration.

The present study contributes to the ongoing debate about the nature of consciousness and metacognition. The ability of LLMs to perform metacognitive-like processes without consciousness challenges traditional views of metacognition as solely a conscious endeavor. This could redefine our understanding of learning and adaptation in both biological and artificial entities, suggesting that metacognition may not necessitate consciousness as traditionally conceived.

**Implications of the study**

*Implications for Generative AI metacognition*

Previous studies have indicated that metacognition may play a key role in developing LLMs as agents (Toy et al. 2024). The present study further provides insight into what aspects of LLM metacognition work well, and what would benefit from further development. As Toy et al. (2024) point out, integrating metacognitive modules allows LLM to become agents which orchestrate different higher order capabilities and adapt their strategies, such as evaluating progress towards goals, explaining reasoning for providing a certain score in evaluation, questions about possible improvements in light of learning, etc. The present study indicates the need for building metacognitive modules that aid reasoning in complex and ambiguous tasks (such as choice of "worst" responses on our test). Previous studies have also pointed that incorporating human-like metacognitive learning abilities could lead to more autonomous AI performance across different tasks (Conway-Smith & West, 2024). The present study, revealed patterns of expert metacognition which may be area of LLM improvements: balance reliance of different set of judgment criteria, rather than rigid adherence to specific frameworks; reflection modules that would call for displays of uncertainty, doubt or elicit other metacognitive comments. Meta-learning techniques appear as a general pathway for moving towards agentic view of LLMs, especially in complex problem solving and decision making under uncertainty. Comparative analysis of metacognition of specific LLMs provides further insight into what specific metacognitive modules may enhance each LLMs performance (Table 9). Integrating different metacognitive functions may be a pathway to more autonomous functioning and emergence of AI as a true metacognitive agent.

*Implications for using Generative AI in coaching and coach training*

Given that all tested LLMs performed better than certified coaches on metacognitive metrics and that qualitative analysis of their judgment of reasoning provided similar deficiency as those demonstrated by certified coaches, it can be argued that AI tools could augment coach training practices. AI could serve as a tool to support humans in their assessment of competencies or providing feedback based on clear criteria. These AI simulators could help users practice skills in realistic scenarios, offering insights based on the user's responses and adapting the coaching strategy accordingly. The study also highlights the need for further development of contextual understanding to handle complex tasks (e.g. ethical dilemmas, ambiguous contexts) effectively.

*Future research*

Key area of future research into LLMs' metacognition are more complex abilities, such as learning from feedback, reassessing performance and adaptation. The present study was based on Zero-shot prompts, making a case for studying metacognition under different feedback prompts, e.g. enhanced Zero-shot prompts incorporating step-by-step thinking, two-shot chain of thought reasoning approaches, or two-shot chain of thought reasoning approaches augmented with step-by-step thinking (Wang et al., 2024). Analogous to the way metacognition develops in children with the role of feedback and from separate domains to its generalized application, we may envision the critical role of human feedback in enhancing metacognition of LLMs (Veenman et al. 2005; Veenman & Spaans, 2005). Finally, integrating longitudinal studies into how AI and human metacognitive abilities evolve with training could offer additional insights into developmental trajectories.

Table 9. Metacognitive profiles of LLMs in context of complex tasks
and suggested metacognitive modules for further development.

| LLM | General metacognitive profile | Best suited tasks | Worst suited tasks | Suggested metacognitive module |
|---|---|---|---|---|
| GPT 4 | Moderate on all metacognitive metrics | Tasks requiring moderate accuracy and sensitivity, such as text summarization or factual question answering. | Complex tasks demanding high discrimination between accurate and inaccurate responses, like medical diagnoses, high stakes decisions. | Adaptive confidence estimation, error pattern recognition. |
| Llama 3 and Mistral Large | Under confidence, low sensitivity, moderate predictive accuracy | Tasks demanding more conservative responses, like quality assurance, preliminary screening in healthcare, customer service triage, simple sentiment analysis etc. | Time-sensitive tasks that require quick decision-making, like real time fraud detection. | Confidence boost, reinforcement training |
| Claude-3-Opus Opus | Low sensitivity, low predictive accuracy, high over confidence | Brainstorming or exploratory text generation tasks where overconfidence isn't a major issue. | Tasks demanding precise accuracy, such as scientific research. | Reducing overconfidence |
| Gemini | Negative sensitivity, low predictive accuracy, moderate overconfidence | Text generation or creative writing tasks where overconfidence is not critical. | High-stakes decision-making tasks requiring accurate calibration and sensitivity, like medical diagnosis | Adaptive confidence estimation to reduce negative sensitivity |

*Limitations of the study*

The present study is exploratory in nature and due to the small sample size of questions and participants, findings are not generalizable. Moreover, due to a small sample size, statistical testing of group differences was not possible, limiting all such claims to descriptive ones. The study was also focused on a very specific type of task performance requiring similar investigation across varieties of contexts. Although the context for analyzing metacognition in the present study was specific (performance on an ICF-mimicking test), previous studies have indicated that some components of metacognitive efficiency are domain general, as well as that training metacognitive efficiency in one domain can enhance it in other domains (Mazancieux et al., 2020). Moreover, tasks of understanding coaching competencies, which were selected for our study of metacognition, can be seen as a relatively complex field of reasoning. Therefore, the present study may be taken as a assessment of metacognitive ability having in mind complexity of first order cognitive task.

## Conclusion

Our findings suggest that LLMs exhibit solid metacognitive performance, demonstrating their potential to monitor complex cognitive tasks with a level of precision that may surpass human metacognition. This underlines the significant role of metacognitive-like processes in the functionality of AI systems, which, despite lacking conscious awareness, appear to manage confidence assessment effectively. However, LLMs lacked the reflective qualities inherent in expert metacognition, particularly in more ambiguous or complex situations.

These findings raise important considerations for the integration of AI in domains that require not only technical competencies but also deep contextual understanding and reasoning. The potential for AI to assist and augment human performance in coaching, decision-making, and learning is vast; however, our research indicates the need for AI systems to evolve beyond rigid algorithmic responses towards more nuanced, context-aware processing that incorporates ethical and reflective capacities.

For future research, this study opens several pathways. It emphasizes the importance of further developing AI's metacognitive modules to enhance their capability to engage in more sophisticated, reflective thinking and ethical reasoning. Exploring the impact of different feedback mechanisms and training protocols on AI's metacognitive abilities could lead to more adaptive and autonomous AI systems. Additionally, longitudinal studies could examine the evolution of AI and human metacognitive capabilities in tandem, offering deeper insights into the interplay of their developmental trajectories.